\documentclass[11pt,twoside,A4]{article} 
\usepackage{times,fancyhdr}
\usepackage[dvips]{graphicx}
\usepackage{latexsym} 
\usepackage[affil-it]{authblk}
\usepackage{amsmath}
\usepackage{amssymb}
\usepackage{setspace}

\usepackage[top=4cm, bottom=4cm, left=4cm, right=4cm]{geometry}

\def\beq{\begin{equation}}
\def\eeq{\end{equation}}
\def\beqn{\begin{eqnarray}}
\def\eeqn{\end{eqnarray}}

\begin{document}
 
\title{Phenomenology of Space-time Imperfection I: Nonlocal Defects}
\author{S.\ Hossenfelder\thanks{hossi@nordita.org}} 
\affil{\small Nordita\\
KTH Royal Institute of Technology and Stockholm University\\
Roslagstullsbacken 23, SE-106 91 Stockholm, Sweden}
\date{}
\maketitle
\begin{abstract}
If space-time is emergent from a fundamentally non-geometric theory it will generically be left with defects. Such
defects need not respect the locality that emerges with the background. Here, we develop a phenomenological
model that parameterizes the effects of nonlocal defects on the propagation of particles. In this model, 
Lorentz-invariance is preserved on the average. We derive constraints on the density of defects from
various experiments.
\end{abstract}

\section{Introduction}

Reductionism has been tremendously successful. It is thus only consequential to question
whether the basic ingredients of today's theories are truly fundamental and to devise tests. To make headway
on a theory of quantum gravity, it has long been suggested the seemingly smooth space-time we experience
might be an emergent feature of an underlying, non-geometric theory that unifies general relativity
with quantum mechanics. While the full theory is still unknown, it is nevertheless possible to investigate
the phenomenology of specific features it is expected to have. Many of these investigations in the
area of quantum gravity phenomenology have been reviewed in \cite{AmelinoCamelia:2008qg,Hossenfelder:2009nu,Hossenfelder:2010zj}. 

We will here focus on the question which observational consequences can be expected 
from space-time imperfections of non-geometric origin. Since Lorentz-invariance violation
is the probably most extensively studied area of Planck-scale physics \cite{Mattingly:2005re,Kostelecky:2008ts}, we will here turn our
attention to the case where Lorentz-invariance is maintained, at least on the average.

If space-time is not fundamental, but emergent from a 
non-geometric theory described by graphs, then we live today in a period in which the graph
approximates to excellent precision a locally connected manifold of dimension four with Lorentzian signature and a metric
tensor whose dynamics is determined by Einstein's field equations. This smooth background however will
still have defects owing to its non-geometric origin. One expects some defects to
prevail simply because in such a scenario perfection begs for additional explanation. This expectation that space-time is left with defects is generic and
independent of the details of the underlying theory making it an ideal opportunity to test quantum
gravity phenomenology. We will in this paper make as few assumption
as possible about the origin of the defects and work towards a parameterization of possible
effects that is as model-independent as possible. 

Unlike defects one normally deals with (eg in crystals), the defects under consideration here occur
on space-time points and do not have world-lines. One can then distinguish
two different types of defects: local ones and nonlocal ones. Local defects
respect the emergent locality in the space-time manifold. A particle that encounters a
local defect will scatter and change direction, but continue its world-line on a continuous 
curve. Nonlocal defects on the other hand do not respect the emergent locality of the 
space-time manifold. A particle that encounters a nonlocal defect continues its path in 
space-time elsewhere, but with the same momentum. 

In principle a space-time defect 
could cause both, a change of position and momentum. But before making things
more complicated by combining these two effects, it seems prudent to first study the simpler cases. In this paper, we will 
develop a model for the nonlocal type of defects. An
accompanying paper \cite{own} deals with the local type of defects. A different model for local space-time defects has 
recently been put forward in \cite{Schreck:2012pf} (see also the discussion in \cite{own}).

The nonlocal defects that we are interested in here cause a particle to discontinuously jump between two points that have
a long distance according to the metric of the approximate manifold\footnote{This can be given meaning to without referral to
the metric of the emergent space-time by instead considering the lengths of closed curves that the two points are part of according
to some distance measure of the underlying structure.}. This distance can potentially be macroscopically large. 
This is not unlike transversable wormholes, except that the connection between the points is not described by geometry itself. 

It has been demonstrated explicitly by
Markopoulou and Smolin in \cite{Markopoulou:2007ha} that spin networks are prone to
develop such defects. This finding, aptly dubbed `disordered locality' in  \cite{Markopoulou:2007ha}, was further studied in \cite{PrescodWeinstein:2009wq} as
to its consequences for the cosmological constant and in \cite{Caravelli:2012wy} for its
modifications of the dispersion relation. The approach in these papers however breaks
Lorentz-invariance by introducing a preferred time-like slicing and is thus not of immediate 
relevance for the following considerations, though it has served as a motivation.

The picture we will investigate here is what modifications a homogeneous space-time distribution of 
nonlocal defects causes, depending on the density of the defects and the probability of a particle to interact with the defect. We will see that
massless particles are stochastically deviated from the lightcone. This deviation can be such 
that the average speed remains the usual speed of light and only stochastic fluctuations around the mean occur.
But there are also cases in which
the photon's average speed can decrease or increase. These modifications become more relevant for long distances and small energies,
in stark contrast to other scenarios where the modifications become relevant for large energies, as in the cases
where Lorentz-symmetry is broken or deformed. In the scenario with nonlocal defects, the photon acquires an effective mass that can, for the case when the
photon's average speed exceeds the usual speed of light, potentially be a tachyonic mass. If we want to include 
the case in which deviations from the lightcone can be spacelike, special attention has to be paid to
causality, which we will discuss in section \ref{disc}.

The effects of nonlocal space-time defects that we consider here are not perturbative in the sense that they 
do not constitute a small deviation from flat space (or some other background metric). A flat space
remains flat in the presence of nonlocal defects. The deviation that the nonlocal defects cause 
can best be understood as a topology change. Space-time defects give rise to deviations that are
small in the sense that we assume the defects are sparse. 

We use the unit convention $\hbar=c=1$. The signature of the metric is $(+,-,-,-)$.

\section{The distribution of defects}
\label{distribution}
 
To develop our model for nonlocal defects, we will assume that the emergent background space-time
is flat Minkowski space, ie we will here not take into account background curvature. Since
violations of Lorentz-invariance have been thoroughly studied, and been ruled out to
high precision already, we will further only consider the case in which Lorentz-invariance is maintained, at least
on the average. We first discuss the classical point particles (section \ref{classical}) and
include quantum uncertainty later (section \ref{quantum}).

The
only presently known probability distribution for points in Minkowski space that preserves
Lorentz-invariance on the average is the result of a Poisson process developed in \cite{Dowker:2003hb, Bombelli:2006nm}.
With this distribution, the probability of finding  N points in a space-time volume $V$ is
\beqn
P_{\rm N}(V) = \frac{(\beta V)^{\rm N} \exp(-\beta V)}{{\rm N}!} \quad, \label{pois}
\eeqn
where $\beta$ is a constant space-time density. 

The 
average value of points that one will find in some volume is the expectation value
of the above distribution and given by
\beqn
\langle {\rm N}(V) \rangle = \sum_{{\rm N}=0}^\infty P_{\rm N}(V) {\rm N} = \beta V \quad.
\eeqn
The variance that quantifies the typical fluctuations around the mean is $\Delta {\rm N} \sim \sqrt{\beta V}$,
and the corresponding fluctuations in the density of points are $\Delta ({\rm N}/V) \sim \sqrt{\beta/V}$. 
In other words, the density fluctuations will be small for large volumes. 

We will
use the distribution (\ref{pois}) to seed the nonlocal defects with an average density $\beta$. In
the following, we will not be concerned with fluctuations in the density as our aim here
is to first get a general understanding for the size of effects caused by nonlocal
defects and using the average will suffice for this purpose. 
The probability is a density over space-time $\beta = L^{d+1}$, where $L$
is a length scale and $d$ is the number of spatial dimensions. With this,
we have introduced a small dimensionless parameter $\epsilon = l_{\rm P}/L$, where
$l_{\rm P}$ is the Planck length, that one expects to determine the size
of effects.

The nonlocal defect causes a jump of the particle to another point. The two points have a non-vanishing distance according
to the Minkowski-metric. In addition to sprinkling the points, we then also have to identify the orientations at the begin and end of the jump,  to 
prevent that directions from being twisted. We will assume that the tangential space at the entry point of
the defect is mapped to the parallel-transported tangential space at the exit point. In 
flat Minkowski space this means simply that the direction of a particle 
is not changed and its momentum stays the same while it seems to jump from one point to
another. 

If the probability distribution of nonlocal defects is constant through space and time, then, to preserve Lorentz-invariance,
we further have to make sure that the distribution of points to which the particle jumps depends
only on Lorentz-invariant quantities. To parameterize this distribution of endpoints we will now consider
one single defect and without loss
of generality assume it is located at $(0,0)$. 

The only four-vectors that we have and that
can be used to construct a Lorentz-invariant distribution are the particle's momentum before
encountering the defect, $p^\nu$, the particle's momentum after encountering the defect, $p'^\nu$, and the space-time distance
between the defect and the point where the particle will continue its path, $y_\nu$. By assumption, the momenta
before and after encountering the defect are identical $p^\nu = p'^\nu$, so we have only two four-vectors left.

The first thing that comes to mind is to distribute the end points for the particle's jump according to the
same Lorentz-invariant distribution as the defects. If space is infinitely extended, this means that the
particle with probability one jumps infinitely far away and essentially disappears. It also means that it gets 
replaced with another particle at a rate depending on the probability of any particle to encounter such a
defect elsewhere. This then effectively describes a local defect, and we will not further consider this case 
here, see however the discussion in \cite{own}.

To describe the case where the endpoint of the jump is determined by Lorentz-invariant quantities
constructed from the particle's trajectory, we will for now focus on the case with 1+1 dimension 
and generalize to higher dimension later. In 1+1 dimensions, $\beta$ has dimensions of one over length squared.
We first consider a massless, right moving particle with momentum $p_\nu p^\nu = 0$, ie $p^\nu = (E,E)$. 

\begin{figure}[h]
\centering \includegraphics[width=7cm]{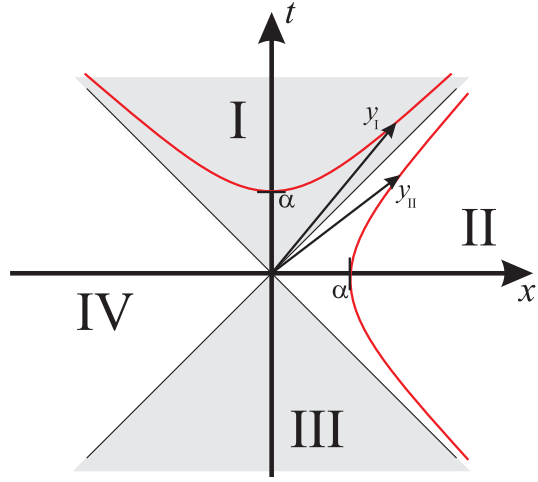}

\caption{Location of $y_I$ and $y_{II}$ in Minkowski space\label{fig1}}
\end{figure}

In 1+1 dimensions, curves of constant distance from a point are hyperbolae in Minkowski space.
The first Lorentz-invariant quantity at our disposal is thus the length of $y_\nu$
\beqn
y_\nu y^\nu = \pm \alpha^2 \quad, \label{req}
\eeqn
where $\alpha$ is a (positive, real) constant of dimension length.
In 1+1 dimensions,  the
hyperbolae in quadrant I (see figure \ref{fig1}) can be parameterized as
\beqn
y_{\rm{I}}^\mu = ( \sqrt{x^2 + \alpha^2}, x) \quad.
\eeqn
Timelike vectors of length $\alpha^2$ ending on the hyperbola in quadrant II can be parameterized as
\beqn
y_{\rm{II}}^\mu = (t, \sqrt{t^2 + \alpha^2}) \quad.
\eeqn
(In the following we write $y$ to refer to both $y_{\rm I}$ and $y_{\rm II}$ together.)

The second Lorentz-invariant quantity that we can construct is the contraction of $p^\nu$ with $y_\nu (\alpha)$.
We parameterize this contraction as 
\beqn
(y_{\rm{I}})_\mu p^\mu =  \alpha \Lambda \quad, \label{reqI}
\eeqn
in quadrant I, and 
\beqn
(y_{\rm{II}})_\mu p^\mu = - \alpha \Lambda  \label{reqII}
\eeqn
in quadrant II. Here, $\Lambda$ is a (real, positive) parameter of dimension mass. For any
given pair of $\Lambda$ and $\alpha$, this singles out a point on the
hyperbola, which is where the particle that hit the defect at $(0,0)$ will
continue its path. Any normalizable function of $\alpha$ and $\Lambda$ will be Lorentz-invariant 
and is thus suitable to parameterize the probability distribution for endpoints of
the particle's jump. The vectors $y_{\rm I}$ and $y_{\rm II}$ are thus functions
of $\alpha$, $\Lambda$ and $p$.

When the particle is deviated from the lightcone to the hyperbola
in quadrant I it will be slowed down. We will thus refer to this case as subluminal. Correspondingly
the case where the jump goes to quadrant II we will refer to as superluminal.

In the superluminal case we can eliminate the explicit dependence of $y_{II}$ on $t$ and
express it through the particle's energy which gives

\beqn
y_{\rm{II}}(\alpha,\Lambda, p) = \frac{\alpha}{2} \left( 
\begin{array}{c} 
E/\Lambda - \Lambda/E \\
E/\Lambda + \Lambda/E
\end{array} \right) \quad. \label{yII}
\eeqn
For the subluminal case one has similarly
\beqn
y_{\rm{I}}(\alpha,\Lambda, p) = \left( 
\begin{array}{cc} 
0 & 1 \\
1 & 0 
\end{array} \right) y_{\rm{II}}(\alpha,\Lambda, p) \quad. \label{yI}
\eeqn

The probability distribution, $P_{\rm{NL}}(\alpha, \Lambda)$, for the endpoints of the
particles' jumps has to be normalizable to unity, $\int P_{\rm{NL}}(\alpha, \Lambda) d\alpha d \Lambda=1$.
We will denote average values with respect to this distribution with brackets that carry the index `NL',
for example
\beqn
\langle y_{\rm I}(p)  \rangle_{\rm NL} = \int y_{\rm I}(\alpha,\Lambda, p) P_{\rm NL}(\alpha, \Lambda)  d\alpha d\Lambda \quad.
\eeqn

We will in the following only roughly quantify the probability distribution by the average values $\langle \alpha \rangle_{\rm NL}$ and 
$\langle \Lambda \rangle_{\rm NL}$ as well as the corresponding variances $\Delta \alpha$ and $\Delta \Lambda$.  The former determine the average distance the particle will jump when it hits the defect,
the latter the width of the distribution around that average value.

In summary, for the particle with momentum $p_\nu$, the requirement Eq. (\ref{reqII}) singles out one
point $y_{\rm{II}}(\alpha,\Lambda, p)$ on the hyperbola to $\alpha$, this is the endpoint of the jump starting in $(0,0)$. The probability that the 
particle makes a jump to this point is then $P_{\rm{NL}}(\alpha, \Lambda)$. This construction is entirely 
Lorentz-invariant. 
The distribution $P_{\rm{NL}}(\alpha,\Lambda)$ parameterizes the effects of the nonlocal defects in causing
a sudden translation. One may say that the defect resembles a beamer rather than a wormhole.

\section{Worldlines in the presence of nonlocal defects}
\label{classical}

We have now parameterized the distribution of nonlocal defects in the density, $\beta$, of the
Poisson-sprinkling, and the typical deviation from the lightcone
that the defect causes, quantified in the average values $\langle \alpha \rangle_{\rm NL}$ and  $\langle \Lambda \rangle_{\rm NL}$ together with the widths $\Delta \alpha$ and $\Delta \Lambda$. These are the free parameters of the model.

From  Eq. (\ref{yII}) we see that the particle is deviated as follows:
\begin{itemize}
\item In the limit $E\gg \Lambda$, $y$ becomes parallel to $(1,1)$, ie the right forward lightcone.
In this limit, one 
jump of the particle skips a distance of $L_{{\rm 1j}} \approx \frac{E \alpha}{ \Lambda}$.
\item In the limit $E\ll \Lambda$, $y_{\rm II}$ becomes parallel to $(1,-1)$, ie the right backward
lightcone in the superluminal case, and $y_{\rm I}$ becomes parallel to $(-1,1)$, ie the left forward lightcone in the subluminal case.
\item In the case $E \approx \Lambda$, $y_{\rm II}$ becomes parallel to $(0,1)$ in the superluminal case,
and $y_{\rm I}$ becomes parallel to $(1,0)$ in the subluminal case.
\end{itemize}
Since we see that the direction of the jump gets closer to the lightcone for
large energies and the deviation becomes noticeable for energies close to $\Lambda$ and below,
we conclude that $\Lambda$ is an modification in the infrared, rather than in the ultraviolet. We
thus expect $\Lambda$ to be a small energy scale. Roughly speaking, $\Lambda$ is the energy at which a
massless particle jumps a distance of $\alpha$, or $\langle \alpha \rangle_{\rm NL}$ on the average.

Note that the particle's momentum, whenever one measures it, remains $(E,E)$. However, the tangential vector of the {\it average} trajectory is no longer parallel to the momentum vector. If we parameterize the particle's
trajectory $x$ with $\tau$, such that $x(0) = 0$, then the trajectory would now 
be $x=\tau (1,1) + \Theta(\tau) y(\alpha, \Lambda, p)$.

This particle can be assigned a pseudo-momentum, $\tilde p$, and pseudo-mass, 
$\tilde m$, that is the four-momentum and mass of a particle that moved straight from the defect to
the endpoint of the jump in the background manifold. In the superluminal case this
pseudo-momentum and pseudo-mass will be that of a tachyon. Requiring that $\tilde p$ be proportional
to $y(\alpha, \Lambda, p)$ but not be a function of $\alpha$ itself, and $\tilde p \to p$ for $\Lambda \to 0$, one finds
\beqn
\tilde p = 2 \frac{\Lambda}{\alpha} y(\alpha,\Lambda, p) \quad,\quad \tilde m = {\rm i} \Lambda \quad,
\eeqn
in the superluminal case.
The speed the particle would have
if it would constantly jump with parameters $\alpha$ and $\Lambda$ is, from Eq. (\ref{yII}),
\beqn
\tilde c_{{\rm sup}} = \frac{E/\Lambda + \Lambda/E}{E/\Lambda - \Lambda/E} \quad.
\eeqn
For $\Lambda/E \ll 1$, the speed of light with the first correction term is $\tilde c_{{\rm sup}} \approx 1 + 2 (\Lambda/E)^2$.
For the subluminal case one finds the same way $\tilde m = \Lambda$, $\tilde c_{{\rm sub}} = 1/\tilde c_{{\rm sup}}$, and
for large energies it is $\tilde c_{{\rm sub}} \approx 1 - 2 (\Lambda/E)^2$.

Now that we have described one jump caused by a nonlocal defect, let us join several. Consider the
photon makes a sequence of $N_{\rm{I}}$ timelike translations with $\alpha_i, i \in{1..N_{\rm{I}}}$, 
and $N_{\rm{II}}$ spacelike translations with $\alpha_j, j \in{1..N_{\rm{II}}}$. Measured 
in the background manifold the trajectory of the photon, instead of having zero proper length, 
will appear to have length $\sum_{i} \alpha_i + \sum_{j} \alpha_j$. The corresponding pseudo-momentum, $\langle \tilde p \rangle$, assigned
to the photon encountering the sequence of defects is the average over the joined jumps.
Thus
\beqn
\langle \tilde p \rangle &=& 2 \frac{\Lambda}{N} \left( \sum_i^{N_{\rm{I}}} \frac{1}{\alpha_i} y_{\rm{I}}(\alpha_i,p,\Lambda) + 
\sum_j^{N_{\rm{II}}} \frac{1}{\alpha_j} y_{\rm{II}}(\alpha_j,p,\Lambda) \right)
 \quad, \\
\langle \tilde m \rangle &=& \sqrt{\langle \tilde p^2 \rangle} \quad,
\eeqn
were $N =N_{\rm{I}} + N_{\rm{II}}$\footnote{Not the same $N$ as in (\ref{pois}).}, and  the sums are taken only over those endpoints in the respective quadrant.
The average speed the photon would appear to have moved with is
\beqn
\langle \tilde c \rangle = \frac{\langle \tilde p_x \rangle}{\langle \tilde p_t \rangle} \quad.
\eeqn
The average speed can be equal to one if there are lightlike and spacelike translations, and will be strictly
larger (smaller) than one if there are only spacelike (timelike) translations. 

The pseudo-momentum is 
proportional to a sum of four-vectors and thus Lorentz-covariant. Thus, the speed assigned to the particle
transforms under the usual Special Relativistic laws. We note that since the pseudo-momentum does not
depend on $\alpha$, the average pseudo-momentum does also not depend on the probability distribution for $\alpha$. 
Inserting
Eqs. (\ref{yI},\ref{yII}), one finds
\beqn
\langle \tilde p \rangle_{\rm NL} &=& E \left( 1 + \frac{\langle \Lambda^2 \rangle_{\rm NL}}{E^2} \frac{N_{\rm{II}} - N_{\rm{I}}}{N_{\rm{I}} + N_{\rm{II}}}, 
1 + \frac{\langle \Lambda^2 \rangle_{\rm NL}}{E^2} \frac{N_{\rm{I}} - N_{\rm{II}}}{N_{\rm{I}} + N_{\rm{II}}}
\right)
 \quad. 
\eeqn
As one expects, if we have a distribution of nonlocal defects that is the same for timelike and spacelike translations ($N_I = N_{II}$) then one
will initially have some fluctuations, but for a large number of jumps $N_{\rm{II}} \approx N_{\rm{I}}$ and so
$\langle \tilde p \rangle_{\rm NL} \to p$. Note that the number of jumps between two nodes on the curve is an invariant. 

In the case $N_I \neq N_{II}$, for example, when jumps happen exclusively into one quadrant, the average of the
path deviates from the lightcone. If $N_I = N_{II} \equiv N $, the average of the path
remains the lightcone, but the variance is still non-vanishing. In other words, even if the average
remains on the lightcone, a representative path will deviate from it. Since
the particle's jumps constitute a random walk in one dimension, we can estimate that the average deviation in this case scales like
\beqn
\Delta y \sim \frac{\sqrt{N}}{E} \Delta (\alpha \Lambda)      \quad. \label{Deltay}
\eeqn
in the direction perpendicular to the direction of propagation, ie $y_{0} - y_{x}$ . Assuming that the variables $\alpha$ and $\Lambda$ are
uncorrelated, we could further rewrite
\beqn
\Delta (\alpha \Lambda) = \sqrt{\langle \alpha \rangle_{\rm NL}^2 (\Delta \Lambda)^2 + \langle \Lambda \rangle_{\rm NL}^2 (\Delta \alpha)^2} \quad. \label{delta}
\eeqn

However, we are not interested in a direct sequence of translations, but in a sequence of translations
interrupted through propagation on the lightcone in the background manifold that respects 
locality. To model this local process we have to understand the background as discretized, even though the
details will not matter in the following. The reason is that a locally finite distribution of points\footnote{Finitely many points in
all causal diamonds.},
as we are using here for the nonlocal defects, has measure zero in a continuous background,
and the probability of any particle's worldline to hit such a defect in a continuous background would be zero.

\begin{figure}[ht]
\includegraphics[width=13cm]{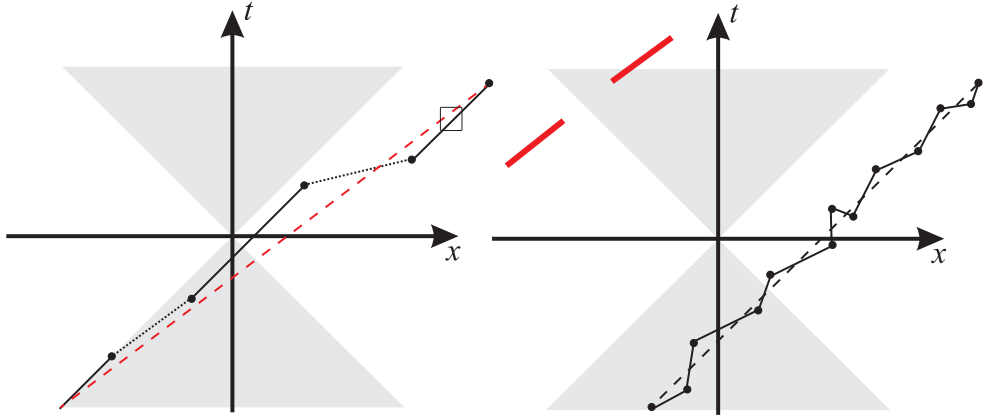}

\caption{Averaging of the particle's trajectory. Left: The solid lines are on the lightcone, the
dotted lines indicate the translations caused by the nonlocal defects. 
The long-dashed (red) line is the effective speed of the particle. Right: 
Zoom in on the box in the left picture. Looking close, it turns out that the particle's
trajectory is a series of short translations (solid) that average to a lightlike curve (dashed). \label{fig2}}
\end{figure}

We will thus consider the local propagation as a sequence of jumps, just that
these are shorter (in the sense that it takes many of them to achieve the same as one nonlocal 
translation). For illustration, see Figure \ref{fig2}. We will here not attempt to develop
an equation of motion for this local process of normal propagation; this is territory of
the underlying theory of quantum gravity. Instead, we just use our existing knowledge and assume 
that the massless particle to good precision moves on the lightcone, and whatever its equation
of motion looks like, it does not deviate too far from it.  It is possible that the fundamental discreteness
has other observable effects besides those caused by defects, but the aim of this
paper is to focus on the nonlocal defects, exactly because evidence for a perfect fundamentally
discrete structure is hard to come by.

However, a concrete example for how such a propagation might look like
has been developed in the Causal Sets approach \cite{Bombelli:1987aa,Dowker:2005tz,Reid:1999gq} in which the dynamical process is described
by `chains' (sequences of points) that are generated iteratively \cite{Ilie:2005qg}. In this case the momentum undergoes
random fluctuations, called `swerves'. These swerves are however unobservably small
if the fundamental length scale of discretization is the Planck length \cite{Mattingly:2007be,Dowker:2010pf}. 

One may have this concrete example in mind, but the details do not matter in the following. The
only thing we need is that the probability to encounter a defect is not zero as it would be in
a continuous background which contains an uncountably infinite amount of points. 

We then describe the motion of the photon as a sequence of $n$ short, local jumps 
with  $\alpha_k, k \in {1..n_{\rm{I}}}$ for quadrant I, $\alpha_l, l \in {1..n_{\rm{II}}}$
for quadrant II, and $n= n_{\rm{I}} + n_{\rm{II}}$. The probability distribution is $P_{\rm{L}}(\alpha,\lambda)$ and the
same in both quadrants.  Here, $\lambda$ plays the
same role for the local jums that $\Lambda$ plays for the nonlocal jumps. The parameters are
the average values taken with the distribution 
$ P_{\rm{L}}(\alpha,\Lambda)$ and its variances. We will assume henceforth that this short-distance structure is at
the Planck-length $\langle \alpha \rangle_{\rm L} \sim 1/\langle \lambda \rangle_{\rm L} \sim  l_{\rm P}$.

In the large $n$ limit with the symmetric distribution of $\alpha$, one then consistently has $\langle \tilde p \rangle = p$.
Now combining both local and nonlocal translations, one gets for the pseudo-momentum
\beqn
\langle \tilde p \rangle &=& \frac{2}{N + n} \left( \sum_i^{N_{\rm{I}}} \frac{\Lambda}{\alpha_i} y_{\rm{I}}(\alpha_i,p,\Lambda) + 
\sum_j^{N_{\rm{II}}} \frac{\Lambda}{\alpha_j} y_{\rm{II}}(\alpha_j,p,\Lambda) \right. \nonumber\\
&+& \left. \sum_i^{n_{\rm{I}}} \frac{\lambda}{\alpha_i} y_{\rm{I}}(\alpha_i,p,\lambda) + 
\sum_j^{n_{\rm{II}}} \frac{\lambda}{\alpha_j} y_{\rm{II}}(\alpha_j,p,\lambda) \right)
 \\
&=& 
E \left( 1 + \frac{\Lambda^2}{E^2} \frac{N_{\rm{II}} - N_{\rm{I}}}{N + n} + \frac{\lambda^2}{E^2} \frac{n_{\rm{II}} - n_{\rm{I}}}{N + n},
1 + \frac{\Lambda^2}{E^2} \frac{N_{\rm{I}} - N_{\rm{II}}}{N+n} + 
\frac{\lambda^2}{E^2} \frac{n_{\rm{I}} - n_{\rm{II}}}{N + n}
\right) \nonumber
\eeqn

In the case $n_I = n_{II} \equiv n$, the average deviation from the light-like path remains the
same as (\ref{Deltay}) because the short jumps do not contribute by assumption.

Let us then turn our attention to the superluminal case where $P_{\rm{NL}} \equiv 0$ in quadrant I, and
take the average of the distribution of nonlocal and local jumps.
We take the limit of an infinitely long curve with $N,n \to \infty$, $n_I - n_{II} \to 0$, while the ratio $N/n := {\cal J} \ll 1$ is held fixed. 
One then gets
\beqn
\langle \tilde p \rangle_{\rm NL} &=& 
E \left( 1 + {\cal J} \frac{\langle \Lambda^2 \rangle_{\rm NL}}{E^2},
1 - {\cal J} \frac{\langle \Lambda^2 \rangle_{\rm NL}}{E^2} 
\right)\quad.
\eeqn
In other words, the particle resembles one with effective mass $\langle \tilde m \rangle = {\rm{i}}  \sqrt{\langle \Lambda^2 \rangle_{\rm NL} {\cal J}}$.
In the subluminal case one gets similarly $\langle \tilde m \rangle = \sqrt{\langle \Lambda^2 \rangle_{\rm NL} {\cal J}}$.

Note however that for the case in which the photon travels only a short distance, it might
be $N=0$, and as long as $N$ is small the effective mass will make discontinuous jumps. The question is
then what means `short distance?'

\section{Particles with spatial width}
\label{quantum}

\begin{figure}[ht]
\includegraphics[width=13.4cm]{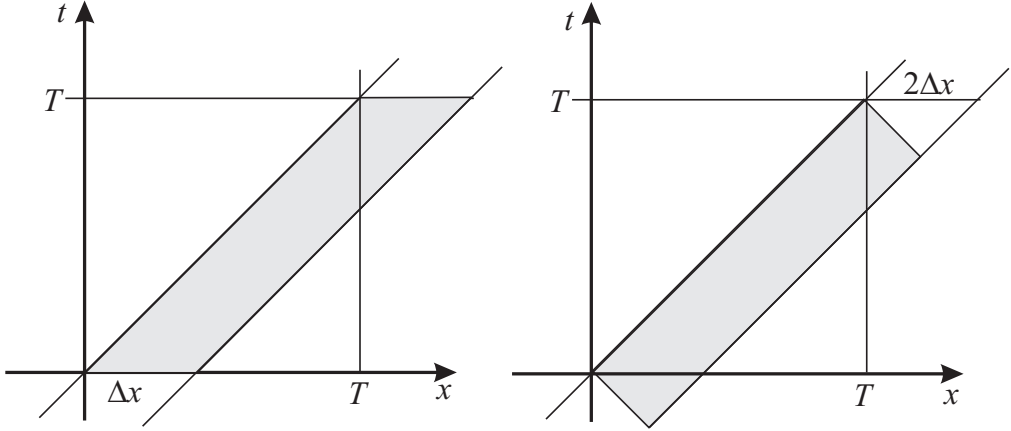}

\caption{To describe quantum particles, we consider worldlines of finite width. While the width and
the propagation time depend on the restframe, the space-time volume that is swept out by the
quantum particle is Lorentz-invariant. \label{fig3}}
\end{figure}

To quantify `short distance,' we leave the image of the
point particle and instead consider a massless particle that covers some
volume in space-time, a natural image when taking into account quantum uncertainty (see Figure \ref{fig3} left).
This particle has a typical width $\Delta x$ and travels for a time $T$.
The width and time depend on the reference frame, but the space-time volume
covered $V = T \Delta x$ does not. Instead of using the spatial
width of the worldline and its temporal extension, we could consider
a causal diamond (see Figure \ref{fig3} right). The former is more
intuitive to think of, the latter is more convenient to analyze Lorentz-
invariance. 

The average number of defects the quantum particle
encounters on its travel is $\beta T \Delta x$.
Now we need an assumption for the interaction probability, ${P}_{\mathrm{int}}$, 
that cannot follow from our previous, entirely kinematic, considerations. We will
assume that the strength of the interaction is determined (as in the
local case, see \cite{own}) by the typical volume of the defect over the
typical volume in which to find the defect. In two dimensions, this means just ${P}_{\mathrm{int}} = \epsilon^2$.

Then in two dimensions the average number of defects the particle interacts with scales as
\beqn
\bar N_2  = {\mathrm P}_{\mathrm{int}}  \beta T \Delta x = \frac{l_{\rm P}^2 T \Delta x}{L^4}  \quad. \label{N}
\eeqn
This means that the average time it takes until the
particle encounters a defect is  
\beqn
\langle T_{\rm{1d}} \rangle \sim \frac{L^4}{l_{\rm P}^2 \Delta x } \quad. \label{ineq}
\eeqn
The limit $T \ll \langle T_{\rm{1d}} \rangle$ is then, for a massless particle, the `short' distance we
are looking for, in which the particle is unlikely to have encountered any defect. 

A word of caution: $\Delta x$ 
is here the length of the baseline of the parallelogram in Figure \ref{fig3}, left. It is only identical to the
width of the particle if the baseline is at constant time, which is not a Lorentz-invariant
statement. This means when interpreting $\Delta x$ in terms of the
width of the wave-function, we best do so in a restframe where the baseline of the parallelogram
is parallel to the $x$-axis, or otherwise first have to derive the relation between both. We will
later, in the section on observational constraints \ref{obs}, identify the width from the
experimental setup.

If  $\langle T_{{\rm 1d}} \rangle \gg T$ 
the particle is unlikely to have made a jump. 
If $T \gg  \langle T_{{\rm 1d}} \rangle$ on the other
hand the particle is likely to have encountered many nonlocal defects already. If that is so, then
the time $T$ that  enters the volume that the particle sweeps out in space-time is no longer the
same as the apparent time $\tilde T$ that one would measure upon arrival, because the volume 
doesn't increase while the particle jumps from one point to the other. 

The relation between the apparent time $\tilde T$ and $T$ is 
\beqn
\tilde T = T \left( \frac{\langle T_{{\rm 1d}} \rangle + \langle T_{{\rm 1j}} \rangle} {\langle T_{\rm 1d} \rangle } \right) \quad, \label{TT}
\eeqn
where $ \langle T_{{\rm 1j}} \rangle_{\rm NL}$ is the average time the particle jumps at each defect. Note that $\tilde T$ could
be zero or negative in the superluminal case. From
Eq. (\ref{yII}) one finds
\beqn
 \langle T_{{\rm 1j}} \rangle_{\rm II} = \frac{1}{2} \left( E \langle \alpha/\Lambda \rangle_{\rm NL} - \frac{\langle \alpha \Lambda \rangle_{\rm NL}}{E} \right) \quad,  \label{TjumpII}
\eeqn
for the superluminal case and
\beqn
 \langle T_{{\rm 1j}} \rangle_{\rm I} = \frac{1}{2} \left(E \langle \alpha /\Lambda\rangle_{\rm NL} + \frac{\langle \alpha \Lambda \rangle_{\rm NL}}{E}\right) \quad \label{TjumpI}
\eeqn
for the subluminal case. The corresponding distances that the particles jump are  $\langle L_{{\rm 1j}} \rangle_{\rm II} =  \langle T_{{\rm 1j}} \rangle_{\rm I}$
for the superluminal case and $ \langle L_{{\rm 1j}} \rangle_{\rm I} = \langle T_{{\rm 1j}} \rangle_{\rm II}$
for the subluminal case. Since the particle is massless $ \langle L_{{\rm 1d}} \rangle = \langle T_{{\rm 1d}} \rangle$.

This means if the measured time $\tilde T$ is large enough so that $T \gg \langle T_{{\rm 1d}} \rangle$, then the particle
jumps over a fraction $\langle T_{{\rm 1j}} \rangle/\langle T_{{\rm 1d}} \rangle$ of this propagation time $\tilde T$ and the
corresponding distance. This can reduce the
probability of it interacting with other particles. We will examine the observational consequences
of this in section \ref{obs}.

For a plane wave, the stochastic deviations of the worldline (\ref{Deltay}) become fluctuations in
the phase of order $\Delta \phi \sim E \Delta y$. This can be calculated from (\ref{yII}), or (\ref{yI}) respectively, and comes
out to be just $\Delta \phi \sim \sqrt{\bar N} \Delta \left( \alpha \Lambda \right)$.
This means that the wave acquires a random phase shift of size $ \Delta (\alpha \Lambda)$ at
each defect, as one could have expected already from (\ref{reqI}) and (\ref{reqII}), and the phase thus
acquires the blurring from the distribution $P_{\rm{NL}}(\alpha, \Lambda)$. It is only the part of
the jump perpendicular to the direction of propagation that blurs the phase. With
the understanding that $\Delta x$ is identified as explained above, we can then express
the random phase shift for a plane wave as 
\beqn
\Delta \phi = \epsilon \sqrt{\frac{T \Delta x}{L^2}} \Delta (\alpha \Lambda) \quad, \label{deltaphi1}
\eeqn
where the variance
can further be rewritten by Eq. (\ref{delta}) so that the phase shift is expressed in the
five parameters we had chosen initially.

We can also use the above considerations to take the limit to classical particles and
to connect to the discussion in the previous section. The average speed
of light is, in the superluminal case, given by $\tilde c = (\langle T_{\rm 1d} \rangle + \langle L_{\rm 1j} \rangle_{\rm II})/(\langle T_{\rm 1d} \rangle + \langle T_{\rm 1j} \rangle_{\rm II})$. In the limit where the deviation is small, $\langle L_{\rm 1j} \rangle_{\rm II}/ \langle T_{\rm 1d} \rangle \ll 1$, one finds $\tilde c \approx 1 + \langle \alpha \Lambda \rangle_{\rm NL}/(E \langle T_{\rm 1d} \rangle)$. With $\Delta x \sim 1/E$, we
can then identify
\beqn
{\cal J} = \epsilon^2 \frac{\langle \alpha \rangle_{\rm NL} }{L}  \frac{1}{\langle \Lambda \rangle_{\rm NL} L}
\frac{\langle \Lambda \rangle_{\rm NL}^2}{\langle \Lambda^2 \rangle_{\rm NL}}
 \quad. \label{J2}
\eeqn

The first factor in this expression is the density of the defects over the density of the
fundamental structure. The second factor is the ratio of the typical distance the particle jumps to
the typical distance of defects. The larger this factor, the more the jumps matter, though we would
expect it to be of order one. The third factor is the ratio of $\hbar$ (set to one) over $\Lambda L$.
It takes into account that when the particle jumps it misses a certain amount of steps of the
fundamental structure, which is why the ratio $N/n$ is not just $l_{\rm P}/L$. The last factor
tells us that the point particle is more strongly affected than a particle with spatial spread if
the distribution is wide, which enters through the difference between the square of the average
and the average of the square. Note that this
identification of ${\cal J}$ however only can be made this way in the limit of small jumps. One could turn these
steps around and start with generalizing ${\cal J}$ for quantum particles, and then obtain
$P_{\rm int}$ from it. 

Of course describing a quantum particle means more than giving a particle a
finite width and a quantum field theoretical description of the nonlocal defects would
be desirable. However, in this present work we aim at first estimating the
effects of nonlocal spacetime defects to see if they are of interest and if they justify
the development of more sophisticated model. 

\subsection{Massless particles in 3+1 dimensions}

For massless particles, the extension of the previously constructed model to
3+1 dimensions, while maintaining Lorentz-invariance, necessitates that we
use additional information about the particle that encounters
the defect. Naively, what we want is that on the average the spatial direction of 
propagation remains unchanged, ie deviations from the direction are uniformly
distributed and do not single out a preferred direction. 

However, to make sense of this requirement (ie
to make sense of the meaning of direction of travel for a massless particle), we
need to take into account the finite spatial width of the photon's wave-function in
two directions perpendicular to the (mean) momentum. For simplicity we will
take the width to be the same in both directions and denote them with $\Delta x_\perp:= \Delta x_1 = \Delta x_2$.
We then assume that these spatial widths remains unchanged when the particle
jumps, ie the mean value
of $y$ in the perpendicular direction $y_\perp (\alpha)$ is zero, $\langle y_\perp \rangle =0$ and $\Delta y_\perp = \Delta x_\perp$.

The need to use additional information about the incident massless
particle goes back to the structure of the Lorentz-group. This is
explained in more detail in section 3.3 of \cite{own}, but can also be
seen as follows.

The reason we cannot introduce jumps into directions transverse to the direction of
propagation without taking into account the width of the photon's wave function is that we 
would have only
two quantities $y^\nu$ and $p_\nu$ at our disposal to construct the endpoint of
the jump. If the total number of dimensions is higher than two, fixing the length
of $y^\nu$ and the product of $y^\nu$ and $p_\nu$ will not single out a point,
but a hypersurface. This hypersurface will include arbitrarily large distances.
Lorentz-invariance would then require a homogeneous probability distribution on
an infinite interval, which is no longer normalizable. We essentially assume that
this non-normalizable distribution arises because a particle with sharply
defined momentum is indeed spatially infinitely extended. Thus taking
into account that in reality the wave function has a finite extension
takes care of the infinite volume.

In 3+1 dimensions we do moreover expect the scaling of the densities to
have different powers than in 1+1 dimensions. The volume that is swept
out by a particle is now $T \Delta x_1 \Delta x_\perp^2$, and the
average number of defects the particle encounters is $T \Delta x_1 \Delta x_\perp^2/L^4$.
$P_{\rm int}$ now scales with the fourth power of epsilon.
This
means that in 3+1 dimensions the average number of defects that a photon moving
in direction $x_1$ interacts with is
\beqn
\bar N \sim \epsilon^4 \frac{T\Delta x_1 \Delta x_\perp^2}{L^4}  \quad, \label{N3}
\eeqn
ie the same as (\ref{N}) except for a prefactor of $\epsilon^2$ and a volume factor stemming from
the two additional perpendicular dimensions. 

Consequently
in 3+1 dimensions the average time the particle travels until encountering
a defect (compare to (\ref{ineq}) in 1+1 dimensions) is
\beqn
\langle T_{\rm{1d}} \rangle \sim \frac{L^8}{l_{\rm P}^4 \Delta x_1 \Delta x_\perp^2} \quad,
\eeqn
 and  the
acquired phase shift is (compare to (\ref{deltaphi1}) in 1+1 dimensions)
\beqn
\Delta \phi \sim \frac{\epsilon^2}{L^2} \sqrt{T\Delta x_1 \Delta x_\perp^2} \Delta (\alpha \Lambda) \quad. \label{deltaphi3}
\eeqn 
Note that the distribution in transverse direction does not contribute to the
phase uncertainty because all points on the hypersurface cut out by $y_\nu p^\nu$ for
a given value of $\alpha$ have the same phase. 

Retracing our earlier steps, we can identify ${\cal J}$ in 3+1 dimensions as (compare to \ref{J2})
\beqn
{\cal J} = \epsilon^4 \frac{\langle \alpha \rangle_{\rm NL}}{L} \frac{1}{\Lambda L}  \frac{\langle \Lambda \rangle_{\rm NL}^2}{\langle \Lambda^2 \rangle_{\rm NL}} \frac{\Delta x_\perp^2}{L^2} \quad,
\eeqn
ie we gained a factor $\epsilon$ per each additional spatial dimension and a factor that compares the 
perpendicular extension to the spacing of defects; the large the perpendicular spread, the more defects
can be hit. This expression is not Lorentz-invariant because taking the limit $E \gg \Lambda$ is not. 

\subsection{Massive particles in 3+1 dimensions}

The notion of a direction transverse to that of propagation is meaningless for massive particles, but for massive
particles one can instead use reference to the particle's restframe. The
requirement that the particle jumps spatially equally far in each spatial direction
leads for a massive particle to a homogeneous distribution on a compact hypersurface. The requirement `equally
far' however is not Lorentz-invariant and valid only in one reference frame. For the massive particle, 
it is natural to choose this reference frame to be the one in which the particle is in rest. Since the
distribution of endpoints $P_{\rm NL}(\alpha,\Lambda)$ depends on the properties of the
ingoing particle, it could be different for massive particles.  

There is however some ambiguity in how to generalize the model to massive particles
because massive particles endow us with an additional scale, the mass of the particle, $m$. 
We could either use this scale instead of $\Lambda$, or use it in addition to $\Lambda$. 
This leads to two different possibilities for massive particles that we will discuss below,
referred to as massive case 1 and massive case 2 respectively.

In
the massive case 1 the parameterization (\ref{reqI}) for the subluminal jumps reads
\beqn
(y_{\rm{I}})_\mu(\alpha,\Lambda, p) p^\mu =  \alpha m \quad, \label{reqIm1}
\eeqn
with (\ref{req}) unchanged. Going into the restframe of the particle, one notes that $y_0 = \alpha$. There
is only one such point on the hyperbola with $\alpha$, that with $y_1=y_2=y_3=0$. Thus,
the particle continues its path in the same direction while being absent for part of its travel, and there
is no spatial deviation. If we apply a boost with 
$\gamma$-factor $E/m$ in $y_1$-direction, the worldline of the massive particle will 
tends towards $\alpha E/m (1,1,0,0)$ and fit with the massless case
when identifying $m$ with the parameter $\Lambda$.

For the superluminal jumps
\beqn
(y_{\rm{II}})_\mu(\alpha, \Lambda, p) p^\mu =  - \alpha m \quad \label{reqIIm1}
\eeqn
leads, in the restframe of the particle, to $y_0 = -\alpha$ and $|\vec y| = \sqrt{2} \alpha$. This too
defines a compact 2-dimensional submanifold, and assigning a uniform probability in the restframe
is possible. This case will however not tend
towards the massless case under large boosts, unless one alters the normalization of $y^2$ to
absorb the $\sqrt{2}$.

In the massive case 2, we can require that in the limit of large boosts, when the
massive particle becomes ultra-relativistic, we find in the direction of the boost the
same average jumps as for the massless particle. This fixes the requirement for the 
subluminal case 2 to
\beqn
(y_{\rm{I}})_\mu(\alpha, \Lambda, p) p^\mu  = m^2 \frac{\alpha}{\Lambda} \label{reqIm2} \quad.
\eeqn
In the particle's restframe, this means $y_0 = m\alpha/\Lambda$ and  $|\vec y| = \alpha \sqrt{(m/\Lambda)^2 -1}$.
Again, this is a compact space on which we can normalize a probability distribution.
We note that the particle can only be translated by the defect if its mass $m \geq \Lambda$. 
The average of the
distribution in the restframe is at $(y_0,0,0,0)$. A boost with $\gamma$-factor $E/m$ in $y_1$-direction then moves the average to $\alpha E/\Lambda (1,1,0,0)$,
ie the same as for the massless particle that moves in $y_1$-direction, provided $m \gg \Lambda$. This determines the factors in (\ref{reqIm2}), though there
could be terms in higher order of $\Lambda/m$.

In the superluminal case 2 we have similarly 
\beqn
(y_{\rm{II}})_\mu(\alpha, \Lambda, p) p^\mu  = - m^2 \frac{\alpha}{\Lambda} \label{reqIIm2} \quad.
\eeqn
In the particle's restframe, this means $y_0 = - m\alpha/\Lambda$ and  $|\vec y| = \alpha \sqrt{(m/\Lambda)^2 + 1}$. 
The average is again at $(y_0,0,0,0)$ and a boost in $y_1$-direction will reproduce on the average the massless case for particle
moving in this direction. As in the case of massless particles, we will assume that the transverse width of
the particle is preserved as it jumps, which means $\Delta y_1 = \Delta y_2 = \Delta y_3 = \Delta x_\perp$, where
$\Delta x_\perp$ is the spatial width of the massive particle in its restframe as it encounters the defect (for
simplicity assumed to be the same in all three directions). In
other words, we assume that the nonlocal defects do not cause additional spatial dispersion.

We will in the
following only consider the massive case 2, where the average longitudinal jump fit well with the massless case in the ultrarelativistic limit.

\section{Observational consequences and constraints}
\label{obs}

In this section we will look at constraints on nonlocal defects from existing experimental data. 
To that end, we will take the so-far anonymous massless particle 
to be a photon. 

There is no direct way to apply to this model existing constraints on the
photon mass because these constraints depend on the way the mass
is generated. In the model discussed here the photon does not actually
acquire a mass. It is just that if the average trajectory of the photon
deviates from the lightcone because of nonlocal defects, then we
can assign an effective mass to the photon, which is the mass that would have caused
the same deviation as the scattering on the defects. So we will have to look for other constraints.

At this point we will reduce the number of parameters in this model 
by assuming that it contains only one new typical length scale, $\langle \alpha \rangle_{\rm{NL}} \sim  \Delta \alpha  \sim L$ in
both quadrants.  We will also assume that there is only one new mass scale and thus $\langle \Lambda \rangle_{\rm NL} \sim \Delta \Lambda$. We are then
left with two parameters, $\langle \Lambda \rangle_{\rm NL}$ and $L$. For the rest of this
section we will drop the brackets and denote the average value with just $\Lambda$. In the following, we further only consider 
the pure superluminal and subluminal cases. The purpose of this section is
to gauge the promise of various observables to roughly constrain the abundance
of nonlocal defects and factors of order one will be omitted. Since we have considered
only the case of a flat background state, the estimates in the following will be good
only so long as the gravitational effects can be neglected. This is a suitable
approximation for Earth based laboratories and for intergalactic propagation 
as long as redshift is negligible.

\subsection{Constraints from single photons in a cavity}

An experiment that delivers good constraints on this model is the tracking of
single photons in an ultrahigh-finesse optical cavity \cite{mirror}. In
this truly amazing experiment, a single photon with a frequency of $\nu \sim$1/mm (in
the deep infrared) is kept bouncing between mirrors that are
approximately 3~cm apart. The photon can be kept a typically
time of 0.5~seconds, which means it travels in total more than
100,000~km back and forth between the mirrors.

\begin{figure}[ht]
\hspace*{-0.8cm}\includegraphics[width=16cm]{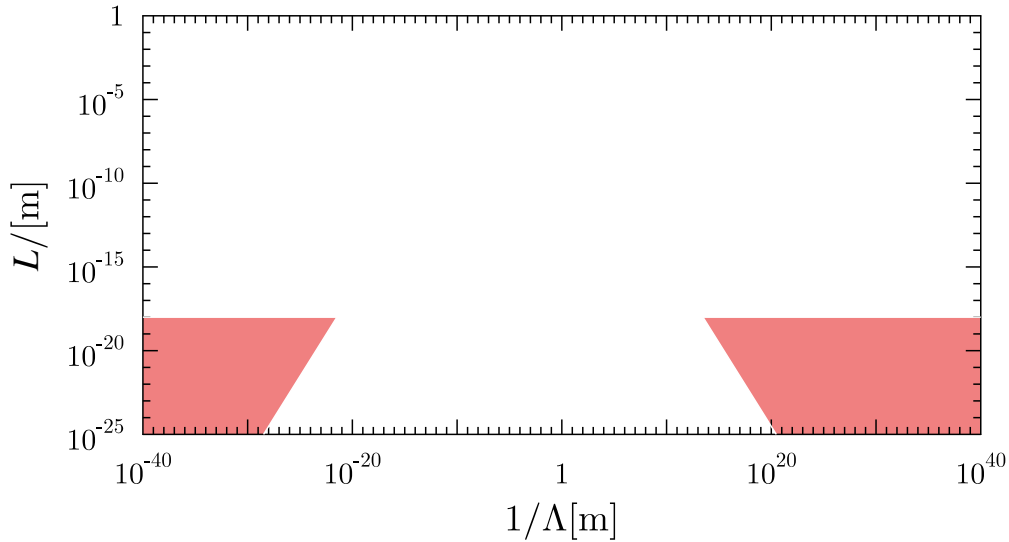}

\caption{Constraints from single photons in a cavity. Shaded regions are excluded. \label{haroche}}
\end{figure}

From this we can derive constraints by noting that a photon of
this energy jumps a distance given by (\ref{TjumpI}) in the superluminal case and (\ref{TjumpII}) in the subluminal
case, and if it jumps then that distance better be less than 3~cm. For $\Delta x_1 \sim 1/\nu$ and
$\Delta x_\perp \sim$~cm we have $T\Delta x_1 \Delta x_\perp^2 \sim 10~\rm{m}^4$, and the allowed parameter range is either
\beqn
L \geq 10^{-18}~{\mathrm{ m}} \quad,
\eeqn
or, if $L \leq  10^{-18}~{\mathrm m}$, then
\beqn
 \frac{L}{2} \Big| \frac{\nu}{\Lambda} \pm \frac{\Lambda}{\nu} \Big| < 3~{\mathrm{cm}} \quad,
\eeqn
where the upper sign stands for the subluminal case and the lower sign for the superluminal case.
These constraints are summarized in Figure \ref{haroche}.

\subsection{Constraints from the GZK cutoff}
\label{gzkbla}

The Pierre Auger Collaboration \cite{Cronin:2007zz,Abreu:2010ab} has found a correlation
between the directions of ultra-high energetic cosmic rays ({\sc UHECR}s), with energies $E_{\rm{GZK}} \gtrsim 6 \times 10^{19}$~eV, 
and active galactic nuclei ({\sc AGN}) up to a distance of about 75 Mpc. This correlation
decreases  when {\sc AGN}s at larger distances are included; beyond 100 Mpc the correlation is
no longer statistically significant. This is evidence for the {\sc GZK} cutoff,
which predicts that the mean free path of protons in the background of the photons
from the cosmic microwave background ({\sc CMB}), is of the order 100 Mpc. Protons of energy high
enough to produce pion pairs when scattering at the {\sc CMB} photons should no longer
reach us. These results from Pierre Auger thus confirm the predictions of the standard model\footnote{A word of caution: 
The statistical significance of these results has decreased with more recent events taken into account. 
It will probably take more time for the situation to become entirely clear.}.

We can use this to derive constraints on the density of nonlocal defects as
follows. If the {\sc UHECR}s protons would jump over a significant part of their path from
the source to earth, then their cross-section with the {\sc CMB} photons would decrease,
and we should have noticed a correlation with {\sc AGN}s over longer distances than predicted by standard
model interactions. This means that
either the protons are unlikely to encounter nonlocal defects or, if they do,
they do not jump very far. 

Deriving a constraint necessitates to find an estimate for the world volume swept
out by the particles we measure. One the one hand the protons are massive particles and experience
dispersion during their travel from the source to earth. Standard quantum mechanical time evolution
leads to an increase of the spatial width of their wave-function. One the other hand, 
if the particle is spread out over a distance much larger than the detector or telescope, one will
only measure the fraction that is collected in the measurement device. 
Increasing the spatial width of the wave-function beyond the size of the
detector means the particle will encounter more defects, but at the same time one is less likely to 
measure anything at all. This would mean that the world
volume of the particles one observes is effectively only $T V_{\rm D}$, where $V_{\rm D}$ is 
the three volume of the detector. 

For the case of the {\sc UHECR}s we can take the detector's
volume to be the volume of the upper atmosphere where the protons scatter, which means
a transverse extension of about $10^{14} \rm{m^2}$. To estimate the width of the proton's wavefunction
upon arrival on earth, we will assume that it was emitted at the {\sc AGN} with
a spatial width $\sigma_0$ determined by the inverse of the temperature of
photoionization equilibrium, which is about 1~eV \cite{Peterson}. In the limit of large times, the
width upon arrival is then $\sigma(T) \sim T/(m \sigma_0)$ in the proton restframe, where
$m\sim$~GeV is the proton restmass. 
Boosting by a $\gamma$-factor of $\sim 10^{10}$, one finds $\Delta x_\perp \sim 10^{5}$~m
and $\Delta x_1 \sim 10^{-5}$~m. Thus, the particles fit well inside the detector and 
the world volume is about $T \Delta x_1 \Delta x_\perp^2 \sim 10^{29} \rm{m}^4$. 

\begin{figure}[ht]
\hspace*{-0.8cm}\includegraphics[width=16cm]{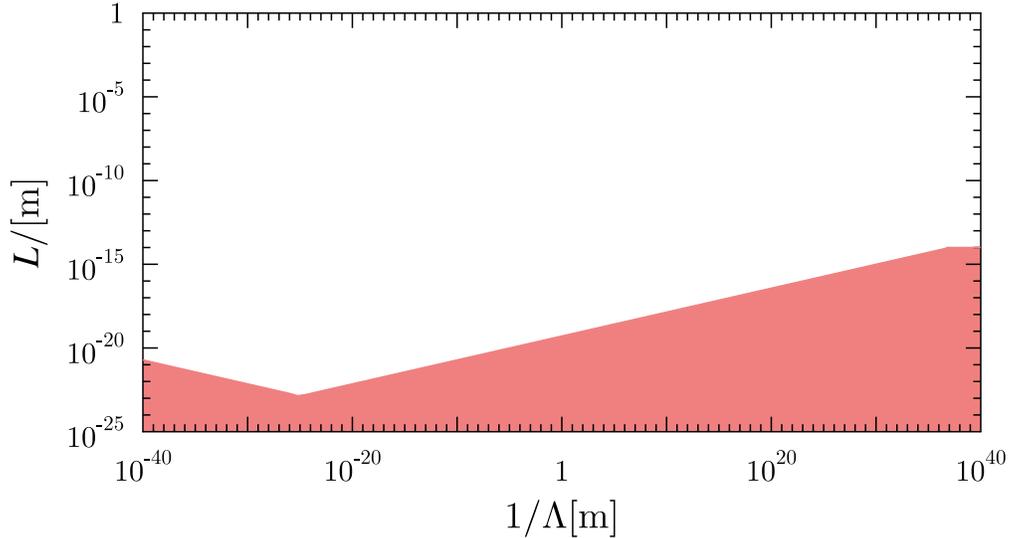}

\caption{Constraints from the correlation of {\sc UHECR}s with {\sc AGN}s. The shaded region is excluded.\label{gzk}}
\end{figure}

This means that either the defects are sparse
\beqn
L \geq 10^{-14}~{\mathrm{ m}} \quad,
\eeqn
or, if $L \leq 10^{-14}$, then $\langle L_{1j} \rangle/\langle L_{1d} \rangle \lesssim 1/10$, where we allow that there is
a 10\% modification to the protons' mean free path, which the correlation analysis by the
Auger collaboration would not have been sensitive to. This leads to the bounds
\beqn
 l_{\rm P}^4 10^{5} {\rm{m}}^3 \Big| \frac{1}{\Lambda} \mp \frac{\Lambda}{E_{\rm{GZK}}^2}\Big| \leq L^7 \quad,
\eeqn
where again the upper sign stands for the subluminal case and the lower sign for the superluminal case. These constraints are summarized in Figure \ref{gzk}. 

\subsection{Constraints from Airy rings of quasars}

Random phase shifts in the propagation of photons have previously been studied
as possible signatures of quantum gravity in the models proposed in  \cite{AmelinoCamelia:1998ax,AmelinoCamelia:1999gg,Ng:1999hm,Christiansen:2005yg}. 
Constraints on these phase shifts can be derived by noting that images from distant
quasars show interference effects, which would be washed out if random phase
shifts were substantial, $\Delta\phi\sim 1$, at that distance and energy \cite{Ragazzoni:2003tn,Steinbring:2006ja}.
It was pointed out already in \cite{Ragazzoni:2003tn} however that modern telescopes allow to set much tighter bounds 
because a telescope focuses a larger part of the light's wavefront than just one wavelength $1/\nu$. 
Since the telescope is sensitive to phase-smearing all over its opening diameter $D$, 
the constraints on loss of Airy rings have more recently been improved to  $\Delta \phi \leq 1/(\nu D) \sim 10^{-8}$ \cite{Tamburini:2011yf}.

These constraints on the model \cite{AmelinoCamelia:1998ax} can also be used for the here discussed propagation in
the presence of nonlocal defects, but first
it is worthwhile to point out the differences. 

In the model going back to \cite{AmelinoCamelia:1998ax},
the phase shifts also accumulate by a random walk, but the number of steps of that walk is proportional
to the distance. Here instead, by the requirement of Lorentz-invariance, it is the space-time volume swept out
by the wavefront that determines the number of steps in the random walk, reflected in equation (\ref{N3}).

Another difference is that the $T$ which determines $\bar N$ is here not necessarily equal to the distance
(time) inferred from the measurement, $\tilde T$. If the particle jumps over part of its path, $T$ is instead only the
distance (time) the photon propagated in the background and generally related to $\tilde T$ by (\ref{TT}). Unlike in the
case of {\sc UHECR} it is difficult to tell how much of its path a photon from a distant quasar might have
omitted due to nonlocal defects. One may speculate that if it jumps over a significant distance, it would experience less
redshift than normally expected. Alas, to take into account background expansion, a more sophisticated
model would first be needed (see also the discussion in section \ref{disc}). The source would also appear less bright because
at any given time, including the one we take the image, a certain fraction of photons is missing from
space-time.
Since this will generally be the case though, regardless of the distance, we lack a reference 
to quantify the relative lack.

Thus, we will consider three cases here. Either the photons don't encounter any defects at all. Or
they do and we have $\langle T_{\rm 1j} \rangle \gg \langle T_{1d} \rangle$ (jumps are short) or
$\langle T_{\rm 1j} \rangle \ll \langle T_{1d} \rangle$ (jumps are long) but too small to blur the interference rings. Assuming that $\Delta x$ is
the coherence time of the light received from the quasar and typically a few wavelengths, we have
 $T/\nu \sim 10^{19}~{\rm m}^2$ for a typical distance of some Gpc and $1/\nu \sim 500$~nm as used in \cite{Tamburini:2011yf}.
The perpendicular width is determined by the opening diameter of the telescope $D\sim$~m (see discussion in section \ref{gzkbla}).

\begin{figure}[ht]
\hspace*{-0.5cm}\includegraphics[width=16cm]{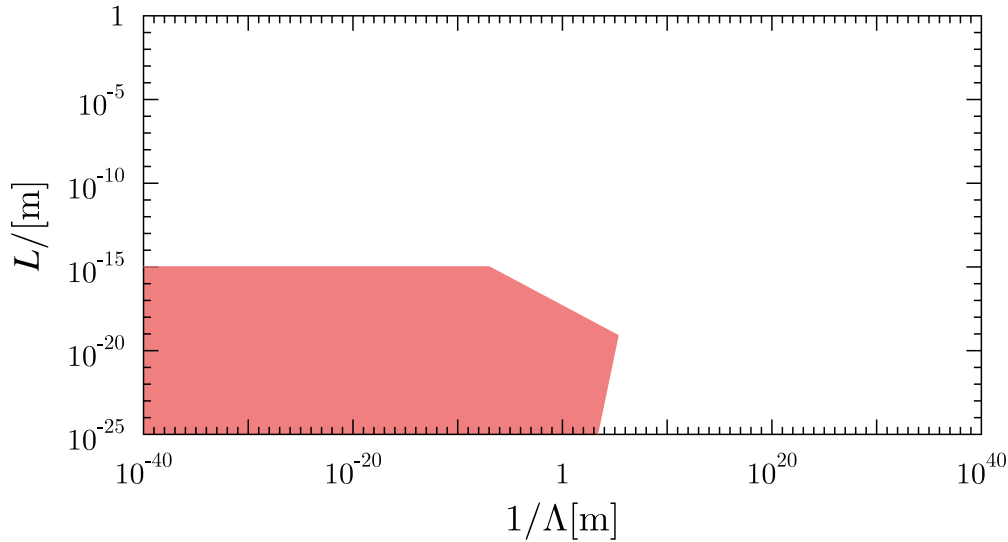}

\caption{Constraints from the observation of interference patters in images of distant quasars. The red shaded region is excluded. \label{quasars}}
\end{figure}

The photons are unlikely to have encountered any defects so long as
\beqn
10^{-15} {\rm m} \leq L \quad.
\eeqn
Jumps are long if $L \leq 10^{-15} {\rm m}$ and
\beqn
l_{\rm P}^4 \frac{D^2}{2 \nu} \Big| \frac{1}{\Lambda} \pm \frac{\Lambda}{\nu^2}\Big| \gg L^7 \quad,
\eeqn
where the upper sign stands for the subluminal case and the lower sign for the superluminal case.
If $L \leq 10^{-15} {\rm m}$ and jumps are short, the requirement that $\Delta \phi \leq 10^{-8}$ leads to
\beqn
10^{-52} ~ {\rm m}^4  \leq L^3 \Lambda^{-1} \quad.
\eeqn
And if the jumps are long one finds
\beqn
10^{-35}~ {\rm m}^{-2}  \Big| \frac{1}{\Lambda^3} \pm \frac{1}{\Lambda^2\nu} \Big| \geq L \quad.
\eeqn
These constraints are summarized in Figure \ref{quasars}.

One can try to improve the constraints on phase shifts by looking for interference rings at longer wavelengths because this will
increase the effective world-volume. The most distant radio pulsar \cite{Momjian:2008vd} is at a
distance of some Gpc and emits at a frequency of $\nu \sim 1/$m. Astrophysical radio signals are measured in 
very large arrays with a baseline of the order $\sim 10^4$~m. Since coherence over the wavefront is necessary
to reconstruct the signal, we can estimate $\Delta \phi \leq 10^{-4}$. However, 
the dependence on the space-time volume enters only with
the square root, so one gains a factor $\sim 10^7$ from the volume, while losing about 
$\sim 10^4$ from the accuracy. This means that the constraint from blurring of
interference rings could potentially be improved with large area telescopes
currently under construction, but the details require further investigation of 
achievable experimental precision and accuracy of image reproduction.

\subsection{Summary of constraints}

\begin{figure}[th]

\hspace*{-0.5cm}\includegraphics[width=16cm]{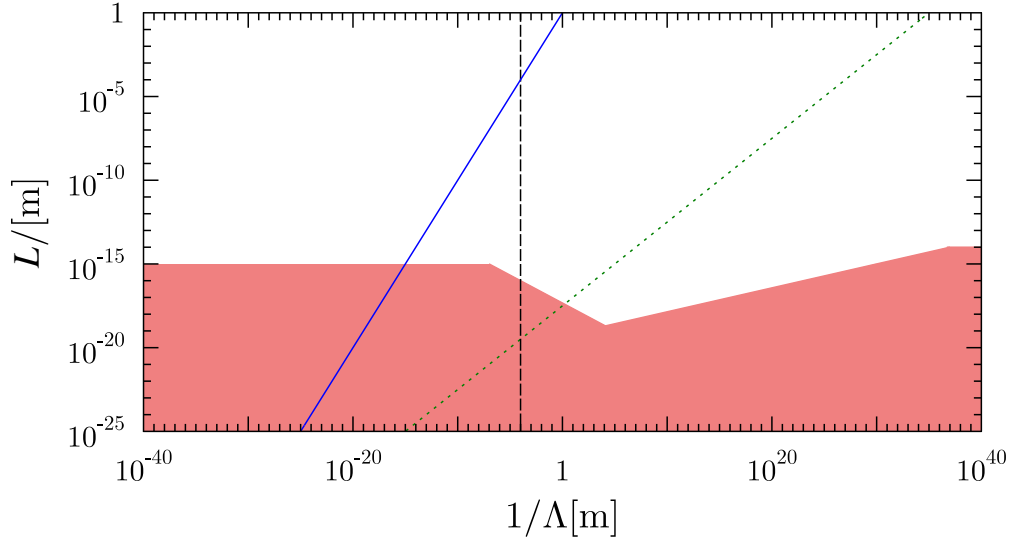}

\caption{Summary of constraints. The red shaded region is excluded. The dashed (black) line
indicates the value of the cosmological constant. The dotted (green)  line depicts the case $\Lambda L =\epsilon$
and the solid (blue) line the case $\Lambda L = 1$. \label{combined}}
\end{figure}

Finally, we can summarize all the above constraints in Figure \ref{combined}. Roughly speaking,
we can take away that nonlocal defects cannot be denser than one per femtometer.  

As noted earlier,
we expect $\Lambda$ to be a scale in the infrared, and so we can speculate that it is of the same order of
magnitude as the cosmological constant. This value is shown in Figure \ref{combined} as
the black, dashed line. The diagonal, dotted (green) line shows the relation $\Lambda L = \epsilon$
and the solid (black) line shows $\Lambda L =1$,
which seem the most natural cases. As we can see the case $\Lambda L = \epsilon$ with $L$ being
comparable to the cosmological constant is not compatible with the data. The relevant 
constraint in this case comes from the blurring of quasar interference rings. 

We see that the constraints considered here are about 10 orders of magnitude too
weak to be sensitive to a density of nonlocal links comparable to the cosmological
constant, which is a natural parameter range in the cosmological setting. However,
the bounds considered here can most likely be improved with a more sophisticated
model which incorporates background curvature, and by exploiting technological
advances in radio astronomy. It thus may be possible to reach the interesting
parameter range. In fact it could be that presently existing data harbors so
far unrecognized evidence
for space-time defects.

\section{Discussion}
\label{disc}

Let us first summarize the assumptions we have made. We have restricted our
examination by requiring Lorentz-invariance to be preserved on the average. 

We have further fixed the typical length
scale for the underlying microscopic discrete structure to be the Planck length,
and later assumed that the typical distance for the nonlocal jumps is similar to the length scale of the 
probability distribution $\langle \alpha \rangle_{\rm NL} = L$. While we did not use a particular
ansatz for the distribution we used its typical width and assumed that the widths are
of the same order of magnitude as the average values $\langle \alpha \rangle_{\rm NL} \sim \Delta \alpha$
and $\langle \Lambda \rangle_{\rm NL} \sim \Delta \Lambda$. These are plausible assumptions, but they could in principle be
relaxed.

As mentioned in section \ref{distribution} we have here not discussed the
possibility that the end points of nonlocal jumps are distributed the same
way as the entry points because this would appear like a local defect.
The difference between the two cases can also be understood as 
the difference between nonlocal connections that can be passed one-way
only and nonlocal connections that can be passed both ways. The case
discussed here is the one-way case. An alternative interpretation of
nonlocal defects that suggests itself is that they cause a translation with
a stochastic element.

Since the nonlocal defects in the superluminal case open the possibility
of particles moving backwards in time in some restframes, 
some words on the issue of causality are in order. The possibility for curves that are
effectively timelike and closed, even though parts of it are jumped over,
necessitates that we require consistency for the time-evolution in order
to prevent causal paradoxa. This is possible, but only seems natural
in the presence of an emergent arrow of time. For such an arrow
of time we have to consider a background filled with quantum fields
of increasing entropy. The Friedmann-Robertson-Walker ({\sc FRW}) background then
provides a preferred slicing according to which one
can require jumps to be `forward'. 

The conservative approach to the causality problem is of course to just restrict
the translations to happen exclusively into quadrant I, which is a Lorentz-invariant
reqirement and does entirely circumvent the issue because then particles
cannot travel over space-like distances, as normally. Note that this restriction
is a restriction on the distribution of translations, ie on $P_{\rm NL}(\alpha)$
and independent of the distribution of start points $\rho_{\rm NL}$.

That having been said, it is clearly desirable to develop this model
to a stage where space-time curvature and quantum fields can be taken into account.
Then it would be possible to use constraints from cosmology, in
particular from the time-evolution of the universe itself. To take
into account a {\sc FRW} background, one needs to make a plausible assumption
for how the density of defects evolves with time. Ideally of course,
this time-dependence will at some point be derived from a
candidate theory for quantum gravity. We hope that the here 
studied case of nonlocal defects in
Minkowski-space serves as a good starting point for such further
developments.

\section{Summary}

We have developed a phenomenological model for nonlocal space-time defects
that maintains Lorentz-invariance on the average.
We parameterized the effects with the space-time density of defects and the
average distance a particle jumps when it encounters a defect, which
depends on the ratio of the particle's
energy over a mass scale in the infrared. The smaller the energy of
the particle, the more pronounced the effect. We then calculated
how the average deviation from the lightcone scales with the
propagation distance and used this to derive constraints on the model from
existing data. We have seen that the nonlocal defects become more
relevant in the infrared for two reasons. First, they create a small effective
mass and second, a particle with large spatial width is more likely to
encounter a defect. 

We found that bounds from various available observations exclude approximately
more than one nonlocal defect in a world volume of a femtometer
to the fourth power. 
These constraints could be improved by studying the interference
patterns of distant radio sources in 
large array telescopes. The vanishing of interference rings could signal
a phase distortion by scattering on nonlocal defects. 
Finally, we noted that it would be desirable
to further develop the model so that it can be applied to a Friedmann-Robertson-Walker
background which would open the possibility to analyze cosmological precision
measurements for possible evidence of space-time defects.

\section*{Acknowledgements}

I thank George Musser and Stefan Scherer for helpful discussions.

\end{document}